\documentclass[onecolumn]{revtex4}%
\usepackage{amssymb}
\usepackage{amsmath}
\usepackage{epsfig}
\usepackage{amsfonts}
\usepackage{graphicx}%
\begin{document}
\title{Vacuum phenomenology of the chiral partner of the nucleon in a linear sigma
model with vector mesons}
\author{Susanna Gallas$^{\text{a}}$, Francesco Giacosa$^{\text{a}}$, and Dirk
H.\ Rischke$^{\text{a,b}}$}
\affiliation{$^{\text{a}}$Institute for Theoretical Physics, Johann Wolfgang Goethe
University, Max-von-Laue-Str.\ 1, D--60438 Frankfurt am Main, Germany}
\affiliation{$^{\text{b}}$Frankfurt Institute for Advanced Studies, Johann Wolfgang Goethe
University, Ruth-Moufang-Str.\ 1, D--60438 Frankfurt am Main, Germany}

\begin{abstract}
We investigate a linear sigma model with global chiral $U(2)_{R}\times
U(2)_{L}$ symmetry. The mesonic degrees of freedom are the standard scalar and
pseudoscalar mesons and the vector and axial-vector mesons. The baryonic
degrees of freedom are the nucleon, $N$, and its chiral partner, $N^{\ast}$,
which is usually identified with $N(1535)$. The chiral partner is incorporated
in the so-called mirror assignment, where the nucleon mass is not solely
generated by the chiral condensate but also by a chirally invariant mass term,
$m_{0}$. The presence of (axial-) vector fields modifies the expressions for
the axial coupling constants of the nucleon, $g_{A}^{N}$, and its partner,
$g_{A}^{N^{\ast}}$. Using experimental data for the decays $N^{\ast
}\rightarrow N\pi$ and $a_{1}\rightarrow\pi\gamma$, as well as lattice results
for $g_{A}^{N^{\ast}}$ we infer that in our model $m_{0}\sim500$ MeV, i.e., an
appreciable amount of the nucleon mass originates from sources other than the
chiral condensate. We test our model by evaluating the decay $N^{\ast
}\rightarrow N\eta$ and the s-wave nucleon-pion scattering lengths
$a_{0}^{(\pm)}$.

\end{abstract}
\maketitle

\section{Introduction}

The theory of the strong interaction, quantum chromodynamics (QCD), has a
global chiral $U(N_{f})_{R} \times U(N_{f})_{L}$ symmetry, for $N_{f}$ flavors
of massless quarks. This symmetry is spontaneously broken in the vacuum, which
has important consequences for hadron phenomenology. Due to confinement of
color charges, all low-energy hadronic properties, such as masses, decay
widths, scattering lengths, etc.\ cannot be inferred from perturbative QCD
calculations. Therefore, effective chiral models are widely used in order to
study the vacuum properties of hadrons. Viable candidates should obey a
well-defined set of low-energy theorems \cite{meissner,gasioro,pisarski}, but
they may still differ in some interesting aspects such as the generation of
the nucleon mass and the behavior at non-zero temperature, $T$, and chemical
potential, $\mu$.

A nucleon mass term $\sim m_{N} \bar{\Psi} \Psi$ explicitly breaks the chiral
$U(N_{f})_{R} \times U(N_{f})_{L}$ symmetry and thus should not occur in a
chiral linear sigma model. Therefore, in the standard linear sigma model of
Refs.\ \cite{gasioro,lee}, the nucleon mass is (mostly) generated by the
chiral condensate, $\langle\overline{q}q\rangle$. (A small contribution also
arises from the explicit breaking of chiral symmetry due to the non-zero
current quark masses.) Similarly, in the framework of QCD sum rules Ioffe
\cite{Ioffe:1981kw} formulated a connection between the quark condensate and
the nucleon mass, now called Ioffe formula: $m_{N}\sim-4\pi^{2} \Lambda
_{B}^{-2} \, \langle\overline{q}q\rangle$, where $\Lambda_{B} \simeq1$ GeV.

However, also other condensates exist, e.g.\ a gluon condensate, and it is not
yet known to what extent they contribute to the nucleon mass
\cite{othersumrules}. This problem can be studied in a chiral model via the
so-called mirror assignment for the chiral partner of the nucleon, which was
first discussed in Ref.\ \cite{lee} and extensively analyzed in
Refs.\ \cite{DeTar:1988kn,jido}. In this assignment, there exists a chirally
invariant mass term $\sim m_{0}$ which does \emph{not} originate from the
quark condensate. The mirror assignment has been subsequently used in
Ref.\ \cite{zschiesche} to study the properties of cold and dense nuclear matter.

In this work we consider a linear sigma model with global chiral
$U(2)_{R}\times U(2)_{L}$ symmetry which includes scalar and pseudoscalar
mesons as well as vector and axial-vector mesons \cite{denis}. We extend this
model by including the nucleon and its chiral partner in the mirror
assignment. The most natural candidate for the chiral partner of the nucleon
is the resonance $N(1535)$ which is the lightest state with the correct
quantum numbers ($J^{P}=\frac{1}{2}^{-}$) listed in the PDG \cite{PDG}. We
also investigate two other possibilities: the well-identified resonance
$N(1650)$ and a speculative, very broad, and not yet discovered resonance with
mass about $1.2$ GeV, which has been proposed in Ref.\ \cite{zschiesche}.

We first study their axial charges which have been the focus of interest in
recent studies of hadron phenomenology [see Ref.\ \cite{Glozman} and
refs.\ therein]. We show that,in the present model, including (axial-) vector
mesons drastically changes the relations of the original model
\cite{DeTar:1988kn}. Without (axial-) vector mesons, $N$ and $N^{\ast}$ have
opposite axial charge, $g_{A}^{N}=-g_{A}^{N^{\ast}}\leq1$. [We remind that, in
the so-called \textquotedblleft naive assignment\textquotedblright, where the
nucleon partner transforms just as the nucleon, one has $g_{A}^{N}%
=g_{A}^{N^{\ast}}=1$ \cite{jido}]. With (axial-) vector mesons, this is no
longer true and we are free to adjust the two axial charges independently,
employing experimental knowledge about $g_{A}^{N}$ and recent lattice QCD data
for $g_{A}^{N^{\ast}}$ \cite{Takahashi}.

Using the decays $N^{\ast}\rightarrow N \pi$ and $a_{1} \rightarrow\pi\gamma$
to determine the other parameters of the model, the mass parameter turns out
to be $m_{0} \sim500$ MeV. This value is between the one derived in
Ref.\ \cite{DeTar:1988kn} and the one from Ref.\ \cite{zschiesche}.

We then test our model studying the decay $N^{\ast}\rightarrow N \eta$ and
pion-nucleon scattering. For $N(1535)$ as chiral partner of the nucleon, the
decay width $N^{\ast}\rightarrow N \eta$ comes out too small, while for
$N(1650)$, it agrees well with experimental data. Pion-nucleon scattering has
been studied in a large variety of approaches [see
Refs.\ \cite{piN,mojzis,ellis,matsui} and refs.\ therein]. Here, we evaluate
the scattering lengths in the framework of the mirror assignment. We find that
the isospin-odd $s$-wave scattering length $a_{0}^{(-)}$ is in good agreement
with experimental data, while the isospin-even scattering length $a_{0}^{(+)}$
strongly depends on the value for the mass of the sigma meson.

Finally, we discuss two possible extensions of our work. The first is an
enlarged mixing scenario. A second pair of chiral partners is added,
e.g.\ $N(1440)$ and $N(1650)$, which also mix with $N(939)$ and $N(1535)$. The
second is the generalization of the chirally invariant mass term $\sim m_{0}$
to a dilatation-invariant mass term. In this case, we argue that $m_{0}$ is a
sum of two contributions, arising from the tetraquark and the gluon
condensates, respectively. The dilatation-invariant mass term also couples a
tetraquark state to the nucleon. We discuss possible implications for nuclear
physics and the behavior of the nucleon mass at non-zero temperature.

This paper is organized as follows. In Sec.\ \ref{II} we present the
Lagrangian of our model and the expressions for the axial charges, the decay
widths $N^{*} \rightarrow N \pi$ and $N^{*} \rightarrow N \eta$, and the
$s$-wave scattering lengths. Section \ref{III} contains our results. In
Sec.\ \ref{IV}, we present a short summary of our work and discuss the two
possible extensions mentioned above, i.e., the enlarged mixing scenario and
the dilatation-invariant mass term. Details of our calculations are relegated
to the Appendices.

Our units are $\hbar= c = 1$, the metric tensor is $g^{\mu\nu} =
\mathrm{diag}(+,-,-,-)$.

\section{The model and its implications}

\label{II}

\subsection{The Lagrangian}

In this section we present the chirally symmetric linear sigma model
considered in this work. It contains scalar, pseudoscalar, vector, and
axial-vector fields, as well as nucleons and their chiral partners including
all globally symmetric terms up to fourth order, see
Ref.\ \cite{denis,Urban:2001ru}. While higher-order terms are in principle
possible, we do not consider them here. In fact, one can argue that they
should be absent in dilation-invariant theories, cf.\ the discussion in
Sec.\ \ref{IV}.

The scalar and pseudoscalar fields are included in the matrix
\begin{equation}
\label{scalars}\Phi= \sum_{a=0}^{3} \phi_{a} t_{a} = (\sigma+i\eta_{N})\,t^{0}
+(\vec{a}_{0}+i\vec{\pi}) \cdot\vec{t}\;,
\end{equation}
where $\vec{t}=\vec{\tau}/2,$ with the vector of Pauli matrices $\vec{\tau}$,
and $t^{0}=\mathbf{1}_{2}/2$. Under the global $U(2)_{R} \times U(2)_{L}$
chiral symmetry, $\Phi$ transforms as $\Phi\rightarrow U_{L} \Phi
U_{R}^{\dagger}$. The vector and axial-vector fields are represented by the
matrices
\begin{subequations}
\label{vectors}%
\begin{align}
V^{\mu}  &  = \sum_{a=0}^{3} V_{a}^{\mu}t_{a} = \omega^{\mu}\, t^{0}
+\vec{\rho}^{\mu} \cdot\vec{t}\; ,\\
A^{\mu}  &  = \sum_{a=0}^{3} A_{a}^{\mu}t_{a} = f_{1}^{\mu} \,t^{0} +\vec
{a}_{1}^{\mu} \cdot\vec{t}\;.
\end{align}
From these fields, we define right- and left-handed vector fields $R^{\mu
}\equiv V^{\mu}- A^{\mu}$, $L^{\mu}\equiv V^{\mu}+ A^{\mu}$. Under global
$U(2)_{R} \times U(2)_{L}$ transformations, these fields behave as $R^{\mu
}\rightarrow U_{R} R^{\mu}U_{R}^{\dagger}\, , \; L^{\mu}\rightarrow U_{L}
L^{\mu}U_{L}^{\dagger}$.

The identification of mesons with particles listed in Ref.\ \cite{PDG} is
straightforward in the pseudoscalar and (axial-) vector sectors, as already
indicated in Eqs.\ (\ref{scalars}), (\ref{vectors}): the fields $\vec{\pi}$
and $\eta_{N}$ correspond to the pion and the $SU(2)$ counterpart of the
$\eta$ meson, $\eta_{N} \equiv(\overline{u}u+\overline{d}d)/\sqrt{2}$, with a
mass of about $700$ MeV. This value can be obtained by ``unmixing'' the
physical $\eta$ and $\eta^{\prime}$ mesons, which also contain $\overline{s}
s$ contributions. The fields $\omega^{\mu}$ and $\vec{\rho}^{\mu}$ represent
the $\omega(782)$ and $\rho(770)$ vector mesons, respectively, and the fields
$f_{1}^{\mu}$ and $\vec{a}_{1}^{\mu}$ represent the $f_{1}(1285)$ and
$a_{1}(1260)$ axial-vector mesons, respectively. (In principle, the physical
$\omega$ and $f_{1}$ states also contain $\overline{s} s$ contributions,
however their admixture is negligible small.)

Unfortunately, the identification of the $\sigma$ and $\vec{a}_{0}$ fields is
controversial, the possibilities being the pairs $\{f_{0}(600),a_{0}(980)\}$
and $\{f_{0}(1370),a_{0}(1450)\}$. In Sec.\ \ref{IVb} a more detailed
discussion of this problem is presented. In the present work, the scalar
assignment affects only the isospin-even $\pi N$ scattering length and we
study its dependence on the sigma mass.

The Lagrangian describing the meson fields reads
\end{subequations}
\begin{align}
\mathcal{L}_{\mathrm{mes}}  &  = \mathrm{Tr}\left[  (D_{\mu}\Phi)^{\dagger
}(D^{\mu}\Phi) -\mu^{2}\Phi^{\dagger}\Phi-\lambda_{2}\left(  \Phi^{\dagger
}\Phi\right)  ^{2}\right]  -\lambda_{1}\left(  \mathrm{Tr}[\Phi^{\dagger}%
\Phi]\right)  ^{2} +c\,(\det\Phi^{\dagger}+\det\Phi)\nonumber\\
&  +h_{0}\,\mathrm{Tr}[(\Phi^{\dagger}+\Phi)] -\frac{1}{4}\mathrm{Tr}\left[
(L^{\mu\nu})^{2}+(R^{\mu\nu})^{2}\right]  +\frac{m_{1}^{2}}{2}\,\mathrm{Tr}
\left[  (L^{\mu})^{2}+(R^{\mu})^{2}\right] \nonumber\\
&  +\frac{h_{1}}{2}\, \mathrm{Tr} \left[  \Phi^{\dagger}\Phi\right]
\mathrm{Tr}\left[  (L^{\mu})^{2}+(R^{\mu})^{2}\right]  +h_{2}\, \mathrm{Tr}%
\left[  \Phi^{\dagger} L_{\mu}L^{\mu}\Phi+\Phi R_{\mu}R^{\mu}\Phi^{\dagger
}\right]  +2h_{3}\, \mathrm{Tr}\left[  \Phi R_{\mu}\Phi^{\dagger} L^{\mu
}\right] \nonumber\\
&  +\mathcal{L}_{3}+\mathcal{L}_{4}\;, \label{meslag}%
\end{align}
where $D^{\mu}\Phi=\partial^{\mu} \Phi+ig_{1}(\Phi R^{\mu}-L^{\mu}\Phi)$, and
$R^{\mu\nu} =\partial^{\mu}R^{\nu}-\partial^{\nu}R^{\mu}$, $L^{\mu\nu
}=\partial^{\mu}L^{\nu}-\partial^{\nu}L^{\mu}$ are the field-strength tensors
of the vector fields. The terms $\mathcal{L}_{3}$ and $\mathcal{L}_{4}$
describe three- and four-particle interactions of the (axial-) vector fields
\cite{denis}, which are not important for this work. We list them in Appendix
\ref{appendixC}. For $c = h_{0} =0$, the Lagrangian $\mathcal{L}%
_{\mathrm{mes}}$ is invariant under global $U(2)_{R} \times U(2)_{L}$
transformations. For $c \neq0$, the $U(1)_{A}$ symmetry, where $A=L-R$, is
explicitly broken, thus parametrizing the $U(1)_{A}$ anomaly of QCD. For
$h_{0} \neq0$, the $U(2)_{R} \times U(2)_{L}$ symmetry is explicitly broken to
the vectorial subgroup $U(2)_{V}$, where $V=L+R$.

The chiral condensate $\varphi= \left\langle 0\left\vert \sigma\right\vert
0\right\rangle =Zf_{\pi}$ emerges upon spontaneous chiral symmetry breaking in
the mesonic sector. The parameter $f_{\pi}=92.4$ MeV is the pion decay
constant and $Z$ is the wavefunction renormalization constant of the
pseudoscalar fields \cite{denis,Strueber}, also related to $\pi$-$a_{1}$
mixing, see Appendix \ref{appendixA} for more details.

We now turn to the baryon sector which involves the baryon doublets $\Psi_{1}$
and $\Psi_{2}$, where $\Psi_{1}$ has positive parity and $\Psi_{2}$ negative
parity. In the mirror assignment they transform as follows:
\begin{equation}
\Psi_{1R}\longrightarrow U_{R}\Psi_{1R}\, ,\; \Psi_{1L}\longrightarrow
U_{L}\Psi_{1L}\, ,\; \Psi_{2R}\longrightarrow U_{L}\Psi_{2R}\, ,\; \Psi
_{2L}\longrightarrow U_{R}\Psi_{2L}\;, \label{mirror}%
\end{equation}
i.e., $\Psi_{2}$ transforms in a ``mirror way'' under chiral transformations
\cite{lee,DeTar:1988kn}. These field transformations allow to write down a
baryonic Lagrangian with a chirally invariant mass term for the fermions,
parametrized by $m_{0}$:
\begin{align}
\mathcal{L}_{\mathrm{bar}}  &  = \overline{\Psi}_{1L}i\gamma_{\mu}D_{1L}^{\mu
}\Psi_{1L} +\overline{\Psi}_{1R}i\gamma_{\mu}D_{1R}^{\mu}\Psi_{1R}
+\overline{\Psi}_{2L}i\gamma_{\mu}D_{2R}^{\mu}\Psi_{2L} +\overline{\Psi}%
_{2R}i\gamma_{\mu}D_{2L}^{\mu}\Psi_{2R}\nonumber\\
&  -\widehat{g}_{1} \left(  \overline{\Psi}_{1L}\Phi\Psi_{1R} +\overline{\Psi
}_{1R}\Phi^{\dagger}\Psi_{1L}\right)  -\widehat{g}_{2} \left(  \overline{\Psi
}_{2L}\Phi^{\dagger}\Psi_{2R} +\overline{\Psi}_{2R}\Phi\Psi_{2L}\right)
\nonumber\\
&  -m_{0}(\overline{\Psi}_{1L}\Psi_{2R} -\overline{\Psi}_{1R}\Psi_{2L}
-\overline{\Psi}_{2L}\Psi_{1R}+\overline{\Psi}_{2R}\Psi_{1L})\;,
\label{nucl lagra}%
\end{align}
where $D_{1R}^{\mu}=\partial^{\mu}-ic_{1}R^{\mu}$, $D_{1L}^{\mu}=\partial
^{\mu}-ic_{1}L^{\mu}$, and $D_{2R}^{\mu}=\partial^{\mu}-ic_{2}R^{\mu}$,
$D_{2L}^{\mu}=\partial^{\mu}-ic_{2}L^{\mu}$ are the covariant derivatives for
the nucleonic fields, with the coupling constants $c_{1}$ and $c_{2}$. (Note
that in the case of local chiral symmetry one has $c_{1}=c_{2}=g_{1}$). The
interaction of the baryonic fields with the scalar and pseudoscalar mesons is
parametrized by $\widehat{g}_{1}$ and $\widehat{g}_{2}$.

The term proportional to $m_{0}$ generates a mixing between the fields
$\Psi_{1}$ and $\Psi_{2}.$ The physical fields $N$ and $N^{\ast}$, referring
to the nucleon and its chiral partner, arise by diagonalizing the
corresponding mass matrix in the Lagrangian (\ref{nucl lagra}):
\begin{equation}
\left(
\begin{array}
[c]{c}%
N\\
N^{\ast}%
\end{array}
\right)  =\frac{1}{\sqrt{2\cosh\delta}}\left(
\begin{array}
[c]{cc}%
e^{\delta/2} & \gamma_{5}e^{-\delta/2}\\
\gamma_{5}e^{-\delta/2} & -e^{\delta/2}%
\end{array}
\right)  \left(
\begin{array}
[c]{c}%
\Psi_{1}\\
\Psi_{2}%
\end{array}
\right)  \; . \label{mixing}%
\end{equation}
The masses of the nucleon and its partner are obtained as:
\begin{equation}
m_{N,N^{\ast}}= \sqrt{ m_{0}^{2} +\left[  \frac{1}{4} (\widehat{g}%
_{1}+\widehat{g}_{2}) \varphi\right]  ^{2}} \pm\frac{1}{4}(\widehat{g}%
_{1}-\widehat{g}_{2})\varphi\;. \label{nuclmasses}%
\end{equation}
The coupling constants $\widehat{g}_{1,2}$ are uniquely determined by the
values of $m_{N}$, $m_{N^{\ast}}$, and the parameter $m_{0}$,
\begin{equation}
\label{g12}\widehat{g}_{1,2}=\frac{1}{\varphi} \left[  \pm(m_{N}-m_{N^{\ast}%
})+\sqrt{(m_{N}+m_{N^{\ast}})^{2} - 4m_{0}^{2}} \right]  \;.
\end{equation}
From Eq.\ (\ref{nuclmasses}) one observes that, in the chirally restored phase
where $\varphi\rightarrow0$, the masses of the nucleon and its partner become
degenerate, $m_{N}=m_{N^{\ast}}=m_{0}$. The mass splitting is generated by
breaking chiral symmetry, $\varphi\neq0$.

Note that the nucleon mass \emph{cannot} be expressed as $m_{N}=m_{0}
+\lambda\,\varphi$, thus $m_{0}$ should not be interpreted as a linear
contribution to the nucleon mass. Such a linearization is only possible in the
case when $m_{0}$ dominates or the chiral condensate dominates. As we shall
see, this does not happen and both quantities are sizable.

The parameter $\delta$ in Eq.\ (\ref{mixing}) is related to the masses and the
parameter $m_{0}$ by the expression:
\begin{equation}
\cosh\delta=\frac{m_{N}+m_{N^{\ast}}}{2m_{0}}.
\end{equation}
When $\delta\rightarrow\infty$, corresponding to $m_{0}\rightarrow0$, there is
no mixing and $N=\Psi_{1}\,, \;N^{\ast}=-\Psi_{2}$. In this case,
$m_{N}=\widehat{g}_{1}\varphi/2$ and $m_{N^{\ast}}=\widehat{g}_{2}\varphi/2$,
thus the nucleon mass is solely generated by the chiral condensate as in the
standard linear sigma model of Refs.\ \cite{gasioro,lee} with the naive
assignment for the baryons.

\subsection{Axial coupling constants}

The expressions for the axial coupling constants of the nucleon and the
partner are derived in Appendix \ref{appendixB}. The result is:
\begin{equation}
\label{axialcoupl}g_{A}^{N}=\frac{1}{2\cosh\delta} \left(  g_{A}%
^{(1)}\,e^{\delta}+ g_{A}^{(2)}\,e^{-\delta} \right)  \;,\;\; g_{A}^{N^{\ast}%
}=\frac{1}{2\cosh\delta} \left(  g_{A}^{(1)}\,e^{-\delta} + g_{A}%
^{(2)}\,e^{\delta}\right)  \;,
\end{equation}
where
\begin{equation}
g_{A}^{(1)}=1-\frac{c_{1}}{g_{1}}\left(  1-\frac{1}{Z^{2}}\right)  ,\;\;
g_{A}^{(2)}=-1+\frac{c_{2}}{g_{1}}\left(  1-\frac{1}{Z^{2}}\right)
\label{gas}%
\end{equation}
are the axial coupling constants of the bare fields $\Psi_{1}$ and $\Psi_{2}$.
At this point, it should be emphasized that the interaction with the (axial-)
vector mesons generates additional contributions to $g_{A}^{N}$ and
$g_{A}^{N^{\ast}}$, proportional to $c_{1}$ and $c_{2}$. We now discuss
several limiting cases, using the fact that $Z$ is required to be larger than
1, cf.\ Eq.\ (\ref{g1}):

\begin{enumerate}
\item[(i)] \emph{Local chiral symmetry:} In this case, the coupling constants
$c_{1}=c_{2}=g_{1}$. This implies $g_{A}^{N}=- g_{A}^{N^{\ast}} =Z^{-2}%
\tanh\delta<1$, which is at odds with the experimental value $g_{A}%
^{N}=1.267\pm0.004$ \cite{PDG}.

\item[(ii)] \emph{Decoupling of vector mesons:} Here, $Z=1$ and $c_{1}%
=c_{2}=0$, and we obtain the results of Ref.\ \cite{DeTar:1988kn}: $g_{A}%
^{N}=-g_{A}^{N^{\ast}}=\tanh\delta$. In the limit $\delta\rightarrow\infty$,
this reduces to $g_{A}^{N}=1$ and $g_{A}^{N^{\ast}}=-1.$ Also in this case the
experimental value for $g_{A}^{N}$ cannot be obtained for any choice of the
parameters. Moreover, a positive value of $g_{A}^{N^{\ast}}$, as found in the
lattice simulation of Ref.\ \cite{Takahashi}, is also impossible.

\item[(iii)] \emph{Decoupling of the chiral partner:} This is achieved in the
limit $\delta\rightarrow\infty$, where $N=\Psi_{1}$ and $N^{\ast}=-\Psi_{2}$.
One has $g_{A}^{N}=g_{A}^{(1)}$ and $g_{A}^{N^{\ast}}=g_{A}^{(2)}$. Since
$Z>1$, it is evident that the ratio $c_{1}/g_{1}$ must be negative in order to
obtain the experimental value $g_{A}^{N}=1.267\pm0.004$ \cite{PDG}.
\end{enumerate}

Note that, in the case of local chiral symmetry, the axial charge of the
nucleon can be also correctly reproduced when introducing dimension-6 terms in
the Lagrangian $\mathcal{L}_{\mathrm{bar}}$,
cf.\ Refs.\ \cite{meissner,gasioro,ellis,Ko,Wilms}, because the coefficients
of these so-called Weinberg-Tomozawa (WT) terms \cite{weinberg,tomozawa} can
be adjusted accordingly. However, such WT terms naturally arise when
integrating out the axial-vector mesons from our Lagrangian, just as in chiral
perturbation theory \cite{mojzis}. In this sense, it would be double-counting
to simultaneously consider axial-vector mesons and WT terms. Our
generalization to a \emph{global\/} chiral symmetry allows a description of
the axial charge without explicitly introducing WT terms.

\subsection{Decay widths}

We now turn to the decays $N^{\ast}\rightarrow N \pi$ and $N^{\ast}\rightarrow
N \eta$. The calculation of the tree-level decay width for $N^{\ast
}\rightarrow N \pi$ from the Lagrangian (\ref{nucl lagra}) is straightforward.
However, the decay $N^{\ast}\rightarrow N\eta$ cannot be directly evaluated
because of the absence of the $s$ quark. In order to proceed, we have to take
into account that
\begin{equation}
\eta=\eta_{N}\cos\phi_{P}+\eta_{S}\sin\phi_{P}\;,
\end{equation}
where $\eta_{N}\equiv(\overline{u}u+\overline{d}d)/\sqrt{2}$, $\eta_{S}
\equiv\overline{s}s$ and $\phi_{P}$ lies between $-32^{\circ}$ and
$-45^{\circ}$ \cite{mixingangle}. Then, the decay amplitude $\mathcal{A}%
_{N^{\ast}\rightarrow N\eta}$ can be expressed as
\begin{equation}
\mathcal{A}_{N^{\ast}\rightarrow N\eta}= \mathcal{A}_{N^{\ast}\rightarrow
N\eta_{N}}\, \cos\phi_{P} +\mathcal{A}_{N^{\ast}\rightarrow N\eta_{S}}\,
\sin\phi_{P}\;.
\end{equation}
In the following, we assume that the OZI-suppressed amplitude $\mathcal{A}%
_{N^{\ast}\rightarrow N\eta_{S}}$ is small, so that to good approximation the
decay width $\Gamma_{N^{\ast}\rightarrow N \eta} \simeq\cos^{2} \phi_{P} \,
\Gamma_{N^{\ast}\rightarrow N \eta_{N}}$. Note that the \emph{physical\/}
$\eta$ meson mass, $m_{\eta}=547$ MeV, enters $\Gamma_{N^{\ast}\rightarrow N
\eta}$. Therefore, also the decay width $\Gamma_{N^{\ast}\rightarrow N
\eta_{N}}$ has to be evaluated for the physical mass $m_{\eta}$, not for
$m_{\eta_{N}}$.

The expression for the decay width $N^{\ast}\rightarrow NP$, where $P=\pi
,\eta$, is (for details, see Appendix \ref{appendixB})
\begin{align}
\Gamma_{N^{\ast}\rightarrow NP}  &  =\lambda_{P}\,\frac{k_{P}}{2\pi}%
\,\frac{m_{N}}{m_{N^{\ast}}}\,\frac{Z^{2}}{32\,\cosh^{2}\delta}\,\left\{
w^{2}\,(c_{1}+c_{2})^{2}\,\left[  (m_{N^{\ast}}^{2}-m_{N}^{2}-m_{P}%
^{2})\,\frac{E_{P}}{m_{N}}+m_{P}^{2}\,\left(  1-\frac{E_{N}}{m_{N}}\right)
\right]  \right. \nonumber\\
&  +\left.  (\widehat{g}_{1}-\widehat{g}_{2})^{2}\,\left(  \frac{E_{N}}{m_{N}%
}+1\right)  +2w\,(\widehat{g}_{1}-\widehat{g}_{2})(c_{1}+c_{2})\,\left(
\frac{m_{N^{\ast}}^{2}-m_{N}^{2}-m_{P}^{2}}{2m_{N}}+E_{P}\right)  \right\}
\;. \label{npidecay}%
\end{align}
The coefficients $\lambda_{\pi}=3$, $\lambda_{\eta}=\cos^{2}\phi_{P}$,
$w\equiv g_{1}\varphi/m_{a_{1}}^{2}$, and the momentum of the pseudoscalar
particle is given by
\begin{equation}
k_{P}=\frac{1}{2m_{N^{\ast}}}\sqrt{(m_{N^{\ast}}^{2}-m_{N}^{2}-m_{P}^{2}%
)^{2}-4\,m_{N}^{2}m_{P}^{2}}\;. \label{kpion}%
\end{equation}
The energies are $E_{P}=\sqrt{k_{P}^{2}+m_{P}^{2}}$ and $E_{N}=\sqrt{k_{P}%
^{2}+m_{N}^{2}}$, because the momenta of the nucleon and the pseudoscalar
particles are equal in the rest frame of $N^{\ast}$.

It is important to stress that, in the mirror assignment, the only way to
obtain a nonzero $N^{\ast}N\pi$ coupling is a nonzero value of the parameter
$m_{0}$. In fact, the coupling is proportional to $\cosh^{-1}\delta\propto
m_{0}$, i.e., when $m_{0}$ increases, also the decay width increases.

In the naive assignment, in which the field $\Psi_{2}$ transforms just like
the field $\Psi_{1}$, a term proportional to $m_{0}$ is not possible, because
it would break chiral symmetry. In this case a mixing term of the form
$\propto\overline{\Psi}_{2}\gamma^{5}\Phi\Psi_{1}+$ h.c.\ is allowed. This
leads to a term $\propto\overline{\Psi}_{2}\gamma^{5} (\sigma+i\gamma^{5}%
\vec{\pi} \cdot\vec{t})\Psi_{1}$, where the pion is coupled to $\Psi_{1}$ and
$\Psi_{2}$ in a chirally symmetric way. However, the very same term also
generates a mixing of $\Psi_{2}$ and $\Psi_{1}$ due to the nonzero vacuum
expectation value of the field $\sigma=\sigma_{0}$. When performing the
diagonalization one obtains two physical fields $N$ and $N^{\ast}$, to be
identified with the nucleon and a negative-parity state such as $N^{\ast
}(1535)$. In terms of the physical fields $N^{\ast}$ and $N$ the coupling
$\overline{N}^{\ast}i\vec{\pi}\cdot\vec{t}N$ vanishes; for the explicit
calculation see Ref.\ \cite{jido}. Thus, in the naive assignment and in the
minimal framework with only one multiplet of scalar and pseudoscalar fields
the decay $N^{\ast}\rightarrow N\pi$ vanishes. One could go beyond this
minimal set-up: a possibility is to include the (axial-) vector mesons into
the Lagrangian of the naive assignment. In this way a nonzero derivative
coupling $\propto\overline{N}^{\ast}\gamma^{\mu}\partial_{\mu}\vec{\pi}%
\cdot\vec{t}N$ survives. A complete study of this scenario, involving also the
scattering lengths, is in preparation.

Alternatively, the inclusion of a second (or more) multiplet(s) of
(pseudo-)scalar mesons, see Refs.\ \cite{cohen,dmh}, coupled to the baryon
fields also leads to a nonvanishing coupling between $N^{\ast}$, the nucleon,
and the pion.

\subsection{$\pi N$ scattering lengths}

\begin{figure}[h]
\begin{center}
\includegraphics[width=15cm]{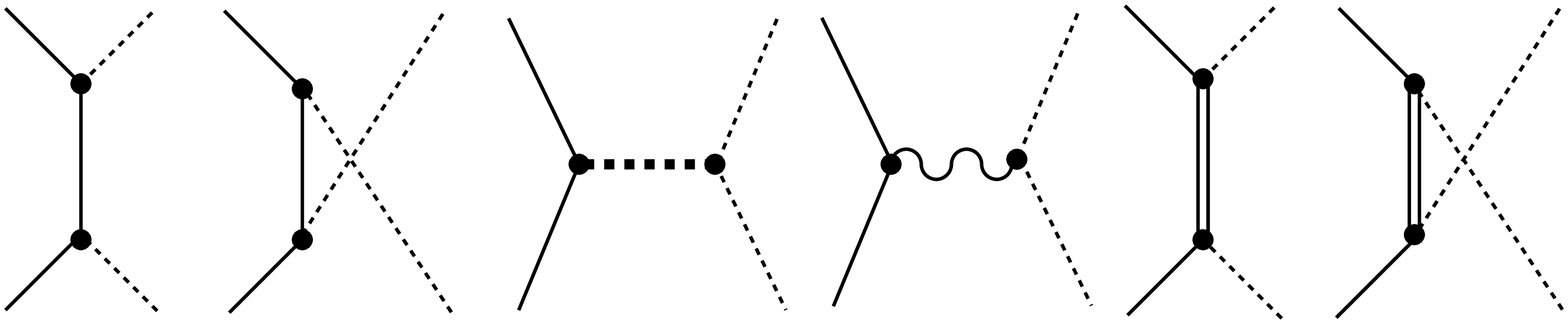}
\end{center}
\caption{Tree-level diagrams contributing to $\pi N$ scattering. Dashed lines
represent the pion, the bold dashed line the $\sigma$ meson, the wavy line the
$\rho$ meson, full lines the nucleon, and double full lines the $N^{\ast}$,
respectively.}%
\label{piN}%
\end{figure}

The general form of the $\pi N$ scattering amplitude is \cite{matsui}:
\begin{equation}
T_{ab}=\left[  A^{(+)}+\frac{1}{2}(q_{1}^{\mu}+q_{2}^{\mu}) \gamma_{\mu
}\,B^{(+)}\right]  \, \delta_{ab} + \left[  A^{(-)}+\frac{1}{2}(q_{1}^{\mu
}+q_{2}^{\mu}) \gamma_{\mu}\,B^{(-)}\right]  \,i\epsilon_{bac}\tau_{c}\;,
\label{Tab}%
\end{equation}
where the subscripts $a$ and $b$ refer to the isospin of the initial and final
states and the superscripts $(+)$ and $(-)$ denote the isospin-even and
isospin-odd amplitudes, respectively. The $\pi N$ scattering amplitudes,
$A^{(\pm)}$ and $B^{(\pm)}$, evaluated from the Lagrangian (\ref{nucl lagra})
at tree-level, involve exchange of $\sigma$ and $\rho$ mesons in the
$t$-channel and intermediate $N$ and $N^{\ast}$ states in the $s$- and
$u$-channels, cf.\ Fig.\ \ref{piN}.

The $s$-wave scattering lengths, $a_{0}^{(\pm)}$, are given by:
\begin{equation}
\label{apm}a_{0}^{(\pm)}=\frac{1}{4\pi(1+m_{\pi}/m_{N})} \, \left(
A_{0}^{(\pm)}+m_{\pi}B_{0}^{(\pm)}\right)  \;,
\end{equation}
where the subscript $0$ at the amplitudes $A^{(\pm)}$, $B^{(\pm)}$ indicates
that they are taken at threshold, i.e., for the following values of the
Mandelstam variables $s,t,u$: $s=(m_{N}+m_{\pi})^{2}$, $t=0$, $u=(m_{N}%
-m_{\pi})^{2}$.

The explicit expression for the isospin-even scattering length can be obtained
from Eq.\ (\ref{apm}) by applying the Feynman rules resulting from the
Lagrangians (\ref{meslag}) and (\ref{nucl lagra}) to the diagrams shown in
Fig.\ \ref{piN}. The result is:
\begin{align}
a_{0}^{(+)}  &  = \frac{1}{4\pi(1+\frac{m_{\pi}}{m_{N}})} \left(  \frac{Z}{2
\cosh\delta}\right)  ^{2} \left(  - \frac{1}{2}\left[  \widehat{g}%
_{1}-\widehat{g}_{2} + \frac{Zf_{\pi}}{2}w(c_{1}+c_{2}) (\widehat{g}%
_{2}-\widehat{g}_{1})\right]  ^{2}\, \frac{(m_{N}+m_{N^{\ast}})(m_{N}%
^{2}+m_{\pi}^{2} -m_{N^{\ast}}^{2})}{(m_{N}^{2}+m_{\pi}^{2}-m_{N^{\ast}}%
^{2})^{2} -4m_{N}^{2}m_{\pi}^{2}} \right. \nonumber\\
&  - w(c_{1}+c_{2})(\widehat{g}_{1} - \widehat{g}_{2}) +\frac{Z f_{\pi}}%
{4}\,(\widehat{g}_{1}-\widehat{g}_{2})\, w^{2}(c_{1}+c_{2})^{2} - w
(c_{1}e^{\delta}-c_{2}e^{-\delta}) (\widehat{g}_{1}e^{\delta}+\widehat{g}%
_{2}e^{-\delta})\nonumber\\
&  + w^{2}m_{N} (c_{1}e^{\delta}-c_{2}e^{-\delta})^{2} +(\widehat{g}%
_{1}e^{\delta}-\widehat{g}_{2}e^{-\delta}) \frac{\cosh\delta}{ Z f_{\pi}}
\left\{  1 + \frac{m_{\pi}^{2}}{m_{\sigma}^{2}}\, \frac{1}{Z^{4}} \left[
Z^{2}-2 +2 (Z^{2} - 1) \left(  1 - \frac{Z^{2}m_{1}^{2} }{m_{a_{1}}^{2}}
\right)  \right]  \right\} \nonumber\\
&  + m_{\pi} \left\{  \left[  \widehat{g}_{1}-\widehat{g}_{2} +\frac{Zf_{\pi}%
}{2}w(c_{1}+c_{2}) (\widehat{g}_{2}-\widehat{g}_{1})\right]  ^{2} \,
\frac{m_{N}m_{\pi}}{(m_{N}^{2}+m_{\pi}^{2} -m_{N^{\ast}}^{2})^{2}-4m_{N}%
^{2}m_{\pi}^{2}} \right. \nonumber\\
&  \left.  \left.  +\, \left[  \widehat{g}_{1}e^{\delta}+\widehat{g}%
_{2}e^{-\delta} -2m_{N}w(c_{1}e^{\delta}-c_{2}e^{-\delta})\right]  ^{2} \,
\frac{m_{N}}{m_{\pi}} \, \frac{1}{m_{\pi}^{2}-4m_{N}^{2}} \right\}  \right)
\; .
\end{align}

Similarly, the expression for the isospin-odd scattering length is given by:
\begin{align}
a_{0}^{(-)}  &  =\frac{1}{4\pi(1+\frac{m_{\pi}}{m_{N}})} \left(  \frac{Z}{2
\cosh\delta}\right)  ^{2} \left(  \left[  \widehat{g}_{1}-\widehat{g}_{2}
+\frac{Zf_{\pi}}{2}w(c_{1}+c_{2}) (\widehat{g}_{2}-\widehat{g}_{1})\right]
^{2} \, \frac{(m_{N}+m_{N^{\ast}})m_{N}m_{\pi}}{(m_{N}^{2} +m_{\pi}%
^{2}-m_{N^{\ast}}^{2})^{2}-4m_{N}^{2}m_{\pi}^{2}} \right. \nonumber\\
&  + \frac{m_{\pi}}{2} \left\{  \left[  \widehat{g}_{1}-\widehat{g}_{2}
+\frac{Z f_{\pi}}{2} w (c_{1}+c_{2})(\widehat{g}_{2} -\widehat{g}_{1})\right]
^{2} \, \frac{m_{N}^{2}+m_{\pi}^{2}-m_{N^{\ast}}^{2}}{ (m_{N}^{2}+m_{\pi}%
^{2}-m_{N^{\ast}}^{2})^{2}-4m_{N}^{2} m_{\pi}^{2}} \right. \nonumber\\
&  - \left[  \widehat{g}_{1}e^{\delta} +\widehat{g}_{2}e^{-\delta}%
-2m_{N}w(c_{1}e^{\delta}-c_{2} e^{-\delta})\right]  ^{2}\, \frac{1}{m_{\pi
}^{2}-4m_{N}^{2}}\nonumber\\
&  - \left.  \left.  w^{2} \left[  (c_{1}+c_{2})^{2}- (c_{1}e^{\delta}%
-c_{2}e^{-\delta})^{2}\right]  +\frac{g_{1}}{m_{\rho}^{2}} \frac{4 \cosh
\delta}{Z^{2}} (c_{1}e^{\delta}-c_{2} e^{-\delta})\right\}  \right)  \; .
\end{align}
Although it is not obvious from these expressions, one can show that the
$s$-wave scattering lengths $a_{0}^{(\pm)}$ vanish in the chiral limit, as
required by low-energy theorems for theories with spontaneously broken chiral symmetry.

\section{Results and discussion}

\label{III}

In this section we present our results. We first discuss the case where the
resonance $N(1535)$ is interpreted as the chiral partner of the nucleon. This
is the most natural assignment because this resonance is the lightest with the
correct quantum numbers. We then consider some important limiting cases.
Finally, we also discuss two different assignments: the resonance $N(1650)$,
which is the next heavier state with the correct quantum numbers listed in
Ref.\ \cite{PDG}, and a \emph{speculative\/} candidate $N(1200)$ with a mass
$M_{N^{\ast}}\sim1200$ MeV and a very large width $\Gamma_{N^{\ast}\rightarrow
N\pi} \gtrsim800$ MeV, such as to have avoided experimental detection up to
now \cite{zschiesche}.

\subsection{$N(1535)$ as partner}

The resonance $N(1535)$ has a mass $m_{N^{\ast}}=(1535 \pm10)$ MeV \cite{PDG}.
The theoretical expressions for $g_{A}^{N}$, $g_{A}^{N^{\ast}}$,
$\Gamma_{N^{\ast}\rightarrow N\pi}$, $\Gamma_{a_{1}\rightarrow\pi\gamma}$
depend on the four parameters $c_{1}$, $c_{2}$, $Z$, and $m_{0}$. Here, $Z$ is
the only parameter entering from the meson sector, see Appendix
\ref{appendixA}.

We determine the parameters $c_{1}$, $c_{2}$, $Z$, and $m_{0}$ by using the
experimental results \cite{PDG} for the decay width $\Gamma_{N^{\ast
}\rightarrow N\pi}^{\exp}=(67.5\pm23.6)$ MeV, the radiative decay of the
$a_{1}(1260)$ meson, $\Gamma_{a_{1}\rightarrow\pi\gamma}^{\exp}=(0.640\pm
0.246)$ MeV, and the axial coupling constant $g_{A}^{N,\exp} =1.267\pm0.004$,
as well as the lattice result $g_{A}^{N^{\ast},\text{latt}}=0.2\pm0.3$
\cite{Takahashi}. With the help of a standard $\chi^{2}$ procedure it is also
possible to determine the errors for the obtained parameters:
\begin{equation}
c_{1}=-3.0\pm0.6 \; , \;\; c_{2}=11.6\pm3.6\; , \;\; Z=1.67\pm0.2\;,
\end{equation}
and
\begin{equation}
m_{0}=(460\pm136)\, \mathrm{MeV}\;.
\end{equation}
The coupling constants $\widehat{g}_{1}$ and $\widehat{g}_{2}$ can be deduced
from Eq.\ (\ref{g12}),
\begin{equation}
\widehat{g}_{1}=11.0\pm1.5\;,\;\; \widehat{g}_{2}=18.8\pm2.4\;.
\end{equation}
The value obtained for $m_{0}$ is larger than the one originally found in
Ref.\ \cite{DeTar:1988kn} and points to a sizable contribution of other
condensates to the nucleon mass. However, because of the non-linear relation
(\ref{nuclmasses}) between the nucleon mass, $m_{0}$, and the chiral
condensate, when switching off $m_{0}$ the nucleon mass is not simply by an
amount $m_{0}$ smaller than the physical value, rather $m_{N} = \widehat
{g}_{1}\varphi/2 \simeq850$ MeV, and thus only slightly smaller than 939 MeV.
The Ioffe formula is thus still approximately justified also in this context.
On the other hand, when varying $\varphi$ from 0 to the physical value
$Zf_{\pi}$, the nucleon mass goes from $m_{0}=460$ MeV to 939 MeV.
Interestingly, the coupling constant $c_{2}$ which parametrizes the
interaction of the nucleon's partner with the (axial-) vector mesons is larger
than the constant $c_{1}$ which parametrizes the interaction of the nucleon
with the (axial-) vector mesons. Nevertheless, when compared with the coupling
$g_{1}\sim6$ (similar in all models with vector mesons and pions) the
constants $c_{1}$ and $c_{2}$ are $\left\vert c_{1}\right\vert \sim g_{1}/2,$
$c_{2}\sim2g_{1}$ i.e., they are related to $g_{1}$ by some numerical factor
of order one. The direct comparison of $c_{1}$ and $c_{2}$ leads to
$\left\vert c_{1}\right\vert \sim c_{2}/4.$

We now test the validity of our model by considering the $\pi N$ scattering
lengths [some preliminary results were already presented in
Ref.\ \cite{Gallas:2009yr}]. The quantity $a_{0}^{(-)}$ depends on $c_{1}$,
$c_{2}$, $Z$, and $m_{0}$, and in addition on $m_{\rho}$ and $g_{1}$. The
latter is a function of $Z$ and $m_{a_{1}}$, cf.\ Eq.\ (\ref{g1}). The values
of $m_{\rho}$ and $m_{a_{1}}$ are known to reasonably good precision
\cite{PDG}, and thus our uncertainty in determining $a_{0}^{(-)}$ is small.
(This will be different for $a_{0}^{(+)}$ which also depends on the poorly
known value of the $\sigma$ meson mass, $m_{\sigma}$.) We obtain
\begin{equation}
a_{0}^{(-)}=(6.04 \pm0.63) \cdot10^{-4}\, \mathrm{MeV}^{-1}\;, \label{a0mst}%
\end{equation}
in agreement with the experimental value measured by the ETH
Z\"urich-Neuchatel-PSI collaboration in pionic hydrogen and deuterium X-ray
experiments \cite{schroder}:
\begin{equation}
a_{0,\exp}^{(-)}=(6.4\pm0.1) \cdot10^{-4}\, \mathrm{MeV}^{-1}\;.
\end{equation}
An even better agreement is expected when including the $\Delta$ resonance
\cite{ellis}.

The scattering length $a_{0}^{(+)}$ depends also on $c_{1}$, $c_{2}$, $Z$, and
$m_{0}$, but in addition on $m_{1}$ and $m_{\sigma}$. The former parametrizes
the contribution to the $\rho$ mass which does not originate from the chiral
condensate: $m_{\rho}^{2}=m_{1}^{2}+\frac{\phi^{2}}{2}(h_{1}+h_{2}+h_{3})$.
Notice that in the present theoretical framework with global chiral symmetry
the KSFR relation \cite{ksfr} is obtained for $m_{1}=0,$ $h_{1}+h_{2}%
+h_{3}=g_{1}^{2}/Z^{2}$. A physically reasonable range of values for $m_{1}$
is between $0$ and $m_{\rho}$. For the lower boundary, the mass of the $\rho$
meson is exclusively generated by chiral symmetry breaking, thus it becomes
massless when $\varphi\rightarrow0$. This is similar to Georgi's vector limit
\cite{Georgivectorlimit} or Brown-Rho scaling \cite{BrownRho}. In principle,
the mass of the $\sigma$ meson varies over a wide range of values; we could
choose $m_{\sigma}\sim0.4$ GeV or $1.37$ GeV, according to the assignment
$f_{0}(600)$ and $f_{0}(1370)$.

\begin{figure}[h]
\begin{center}
\includegraphics[scale = 0.80]{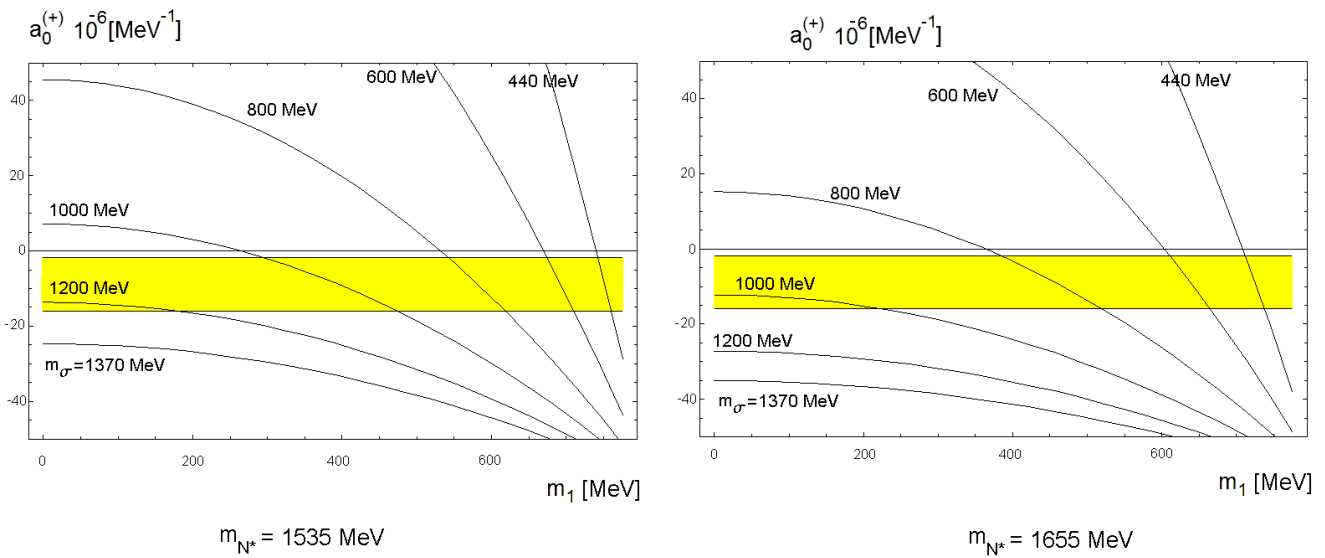}
\end{center}
\caption{The isospin-even scattering length $a_{0}^{(+)}$ as a function of
$m_{1}$ for fixed values of $m_{\sigma}$, for the assignment $N^{\ast
}=N(1535)$ (left panel) and $N^{\ast}=N(1650)$ (right panel). The experimental
range is shown by the grey (online: yellow) band.}%
\label{a0plus}%
\end{figure}

Since the allowed range of values for $m_{1}$ and $m_{\sigma}$ is large, we
choose to plot the scattering length $a_{0}^{(+)}$ as function of $m_{1}$ for
different choices of $m_{\sigma}$; the result is shown in Fig.\ \ref{a0plus}.
The experimental result \cite{schroder}
\begin{equation}
a_{0,\exp}^{(+)}=(-8.8\pm7.2) \cdot10^{-6}\; \mathrm{MeV}^{-1}%
\end{equation}
is shown as grey (online: yellow) band. One observes that for small values of
$m_{\sigma}$ one requires a large value for $m_{1}$ in order to reproduce
experimental data. For increasing $m_{\sigma}$, the required values for
$m_{1}$ decrease. For $m_{\sigma} \agt 1.37$ GeV, $a_{0,\exp}^{(+)}$ cannot be
reproduced for any value of $m_{1}.$ This, however, does not exclude a heavy
$\sigma$ meson, rather, it indicates that an additional light scalar-isoscalar
resonance needs to be included as discussed in Sec.\ \ref{IVb}.

For the decay $N^{\ast}\rightarrow N\eta$, we obtain with Eq.\ (\ref{npidecay}%
) the result
\begin{equation}
\Gamma_{N^{\ast}\rightarrow N\eta}= \left(  10.9 \pm3.8\right)  \,
\mathrm{MeV}\;,
\end{equation}
where the error also takes into account the uncertainty in the pseudoscalar
mixing angle $\phi_{P}=-38.7^{\circ} \pm6^{\circ}$. We observe that
$\Gamma_{N^{\ast}\rightarrow N\eta}$ is about a factor 7 smaller than
$\Gamma_{N^{\ast}\rightarrow N\pi}$, which is in reasonable agreement with the
naive expectation based on the relation $\lambda_{\eta}/\lambda_{\pi}=
\cos^{2} \phi_{P}/3 \simeq0.097$. However, it is clearly smaller than the
experimental value $\Gamma_{N^{\ast}\rightarrow N\eta}^{\exp}=(78.7 \pm24.3)$
MeV \cite{PDG}. The agreement could be improved if one generalizes our
discussion to the $SU(3)$ case and includes a large OZI-violating
contribution, or if one considers an enlarged mixing scenario as discussed in
Sec.\ \ref{IVa}.

\subsection{Limiting cases}

We now consider three important limiting cases. In all of these $N(1535)$ is
taken as chiral partner of the nucleon.

\begin{enumerate}
\item[(i)] \emph{Local chiral symmetry:} This case is obtained by setting
$g_{1}=c_{1}=c_{2}$ and $h_{1}=h_{2}=h_{3}=0$. As a consequence, $m_{\rho
}=m_{1}$, $m_{a_{1}}^{2} = m_{\rho}^{2} + (g_{1} \varphi)^{2}$, $Z=m_{a_{1}%
}/m_{\rho}$. Using the experimental values for $\Gamma_{N^{\ast}\rightarrow
N\pi}$ and $\Gamma_{a_{1}\rightarrow\gamma\pi}$ one obtains
\begin{equation}
m_{0}=\left(  730 \pm229\right)  \, \mathrm{MeV}\;.
\end{equation}
As a consequence, $g_{A}^{N}=-g_{A}^{N^{\ast}} \equiv Z^{-2} \tanh\delta= 0.33
\pm0.02$, both at odds with experimental and lattice data. The scattering
length $a_{0}^{(-)}$ is in the range of the experimental data, $a_{0}^{(-)}=
(4.9 \pm1.7) \cdot10^{-4}$ MeV$^{-1}$. Since $m_{1}= m_{\rho}$ is fixed, the
isospin-even scattering length only depends on $m_{\sigma}$. Thus, for a given
value of $m_{\sigma}$, we obtain a single value with theoretical errors:
$a_{0}^{(+)}= \left(  7.06 \pm3.12\right)  \cdot10^{-6}$ MeV$^{-1}$ for
$m_{\sigma}=1.37$ GeV and $a_{0}^{(+)}=\left(  4.46 \pm0.11\right)
\cdot10^{-5}$ MeV$^{-1}$ for $m_{\sigma}=0.44$ GeV, which is outside the range
of the experimental error band. As already argued in Refs.\ \cite{denis,Wilms}
we conclude that the case of local chiral symmetry (in the present model
without higher-order terms) is not capable of properly reproducing low-energy phenomenology.

\item[(ii)] \emph{Decoupling of vector mesons:} This corresponds to
$g_{1}=c_{1}=c_{2}=h_{1}=h_{2}=h_{3}=0$, and thus $Z=1$ and $w=0$. Using the
decay width $\Gamma_{N^{\ast}\rightarrow N\pi}=(67.5\pm23.6)$ MeV one obtains
\begin{equation}
m_{0}=\left(  262\pm46\right)  \,\mathrm{MeV}\;,
\end{equation}
in agreement with Ref.\ \cite{DeTar:1988kn}. As a result $g_{A}^{N}%
=-g_{A}^{N^{\ast}}=0.97\pm0.01$, in disagreement with both experimental and
lattice data. The description of the scattering lengths also becomes worse;
the isospin-odd scattering length $a_{0}^{(-)}=(5.7\pm0.47)\cdot10^{-4}$
MeV$^{-1}$, which is just outside the experimental error band. Also in this
case, the isospin-even scattering length assumes a single value (with
theoretical errors) for given $m_{\sigma}$: $a_{0}^{(+)}=\left(
1.08\pm0.05\right)  \cdot10^{-4}$ MeV$^{-1}$ for $m_{\sigma}=1.37$ GeV and
$a_{0}^{(+)}=\left(  -7.55\pm0.19\right)  \cdot10^{-4}$ MeV$^{-1}$ for
$m_{\sigma}=0.44$ GeV, i.e., two orders of magnitude away from the
experimental value. We thus conclude that vector mesons cannot be omitted for
a correct description of pion-nucleon scattering lengths.

\item[(iii)] \emph{Decoupling of the chiral partner:} This is obtained by
sending $m_{0}\rightarrow0$ or $\delta\rightarrow\infty$. The partner
decouples and we are left with a linear $\sigma$ model with vector and
axial-vector mesons. The decay $\Gamma_{N^{\ast}\rightarrow N\pi}$ vanishes in
this case and is obviously at odds with the experiment. Using the experimental
values for $g_{A}^{N}$ and $\Gamma_{a_{1}\rightarrow\gamma\pi}$ to fix the
parameters $c_{1}$ and $Z$ ($c_{2}$ and $\widehat{g}_{2}$ play no role here
because of the decoupling of the partner), we obtain
\begin{equation}
c_{1}=-2.59\pm0.51\;,\;\;Z=1.66\pm0.2\;.
\end{equation}
The scattering lengths are:
\begin{equation}
a_{0}^{(-)}=(5.99\pm0.66)\cdot10^{-4}\,\mathrm{MeV}^{-1}\;,
\end{equation}
and $a_{0}^{(+)}$ shows a similar behavior as shown in Fig.\ \ref{a0plus}. We
conclude that the role of the partner is marginal in improving the scattering
lengths. It could be omitted, unless one wants to consider, in the framework
of the mirror assignment, its decay into nucleon and pseudoscalar particles.
\end{enumerate}

\subsection{Other candidates}

In this subsection, we discuss two more exotic possibilities for the chiral
partner of the nucleon.

\begin{enumerate}
\item[(i)] \emph{$N(1650)$ as partner.} The resonance $N(1650)$ has a mass
$m_{N^{\ast}}=(1655\pm15)$ MeV and a decay width $\Gamma_{N^{\ast}\rightarrow
N\pi}^{\exp}=(128\pm44)$ MeV \cite{PDG}. The axial coupling constant measured
in the lattice simulation of Ref.\ \cite{Takahashi} reads $g_{A}^{N^{\ast
},\text{latt}}=0.55\pm0.2$. By following the previous steps we obtain
\begin{equation}
c_{1}=-3.3\pm0.7\;,\;\; c_{2}=14.8\pm3.4\;,\;\;Z=1.67\pm0.2\;,
\end{equation}
and
\begin{equation}
m_{0}=\left(  709\pm157\right)  \, \mathrm{MeV}\;.
\end{equation}
This leads to the coupling constants
\begin{equation}
\widehat{g}_{1}=9.45\pm1.81\;,\;\; \widehat{g}_{2}=18.68\pm2.68\;.
\end{equation}
In this case $m_{0}$ is even larger than before. However, as in the case of
$N(1535)$, the quantity $\widehat{g}_{1}\varphi/2\simeq730$ MeV is still
sizable and similar to $m_{0}$. The result for the isospin-odd scattering
length
\begin{equation}
a_{0}^{(-)}=(5.90\pm0.46) \cdot10^{-4} \,\mathrm{MeV}^{-1}%
\end{equation}
is similar to the case of $N(1535)$. The isospin-even scattering lengths
behave similarly as before, cf.\ Fig.\ \ref{a0plus}, however, slightly smaller
values for $m_{\sigma}$ are required in order to reproduce the experimental data.

The decay width into $N\eta$ is $\Gamma_{N^{\ast}\rightarrow N\eta}%
=(18.3\pm8.5)$ MeV, which should be compared to $\Gamma_{N^{\ast}\rightarrow
N\eta}^{\mathrm{exp}}=(10.7\pm6.7)$ MeV. Thus, in this case the decay width is
in agreement with the experimental value. However, we then face the problem of
how to describe the $N(1535)$ resonance, cf.\ the discussion in
Sec.\ \ref{IVa}.

\item[(ii)] \emph{Speculative candidate $N(1200)$ as partner.} We consider a
\emph{speculative\/} candidate $N(1200)$ with a mass $m_{N^{\ast}}\sim1200$
MeV and a very large width $\Gamma_{N^{\ast}\rightarrow N\pi}\agt 800$ MeV,
such as to have avoided experimental detection up to now. The reason for its
introduction was motivated by properties of nuclear matter \cite{zschiesche}
and further on investigated in Ref.\ \cite{dexheimer} in the context of
asymmetric nuclear matter present in a neutron star. Regardless of the precise
value of the axial coupling constant of the partner (which is unknown for this
hypothetical resonance) one obtains $m_{0}>1$ GeV. This, in turn, implies a
large interaction of $N$ and $N^{\ast}$. As a consequence, both scattering
lengths turn out to be off by two order of magnitudes: $a_{0}^{(-)}\sim
10^{-2}$ MeV$^{-1}$ and $a_{0}^{(+)}\sim10^{-4}$ MeV$^{-1}$. Thus, we are led
to \emph{discard\/} the possibility that a hypothetical, not yet discovered
$N(1200)$ exists.
\end{enumerate}

\section{Summary and outlook}

\label{IV}

In this paper, we investigated a linear sigma model with global chiral
$U(2)_{R}\times U(2)_{L}$ symmetry, where the mesonic degrees of freedom are
the standard scalar and pseudoscalar mesons and the vector and axial-vector
mesons. In addition to the mesons, we included baryonic degrees of freedom,
namely the nucleon and its chiral partner, which is incorporated in the model
in the so-called mirror assignment.

We used this model to study the origin of the mass of the nucleon, the
assignment and decay properties of its chiral partner and the pion-nucleon
scattering lengths. The mass of the nucleon results as an interplay of the
chiral condensate and a chirally invariant baryonic mass term, proportional to
the parameter $m_{0}$. When the chiral partner of the nucleon is identified
with the resonance $N^{\ast}\equiv N(1535)$, the parameter $m_{0}\simeq500$
MeV is obtained as a result of a fitting procedure which involves the three
experimentally measured quantities $N^{\ast}\rightarrow N\pi,$ $a_{1}%
\rightarrow\pi\gamma$, $g_{A}^{N}$, and the quantity $g_{A}^{N^{\ast}}$
evaluated on the lattice. The isospin-odd scattering length $a_{0}^{(-)}$ is
then fixed and found to be in good agreement with experimental data. The
isospin-even scattering length depends, in addition, strongly on the mass of
the $\sigma$ meson, see Fig.\ \ref{a0plus} and the discussion in
Sec.\ \ref{IVb}. The decay width $N^{\ast}\rightarrow N\eta$ turns out to be a
factor of eight smaller than the experimental value.

The obtained value $m_{0}\simeq500$ MeV implies that a sizable amount of the
nucleon mass does not originate from the chiral condensate. As this result is
subject to the assumptions and the validity of the employed chiral model, most
notably due to the identification of the chiral partner with $N(1535)$ and to
the mathematical properties of the mirror assignment, future studies of other
scenarios, incorporating new results both from the experiment and the lattice,
are necessary to further clarify this important issue of hadron physics.

It should also be noted that the results presented in this work are based on a
tree-level calculation. The inclusion of loops represents a task for the
future. Nevertheless, we expect that the results will not change
qualitatively: On the one hand, while the dimensionless couplings of the model
$g_{1}$, $c_{1},$ $c_{2},$ $\widehat{g}_{1},$ $\widehat{g}_{2}$ are large, the
contribution of loops is suppressed according to large--$N_{c}$ arguments
\cite{largenc}. On the other hand, in our model we have included from the very
beginning the relevant resonances which contribute as virtual states to
processes, thus reducing the effects of loops in the model. To clarify the
latter point, consider the $\rho$ meson exchange in $\pi N$ scattering. In an
approach in which the $\rho$ meson is not directly included, its contribution
could only be obtained after a corresponding loop resummation, while in our
approach it is taken directly into account by a tree-level exchange diagram.

We studied three important limiting cases: (i) In the framework of local
chiral symmetry it is not possible to correctly reproduce low-energy
phenomenology. (ii) It is not admissible to neglect (axial-) vector mesons.
They are crucial in order to obtain a correct description of the axial
coupling constants and $\pi N$ scattering lengths. (iii) The role of the
partner $N^{\ast}$ has only a marginal influence on the scattering lengths.

We have also tested other assignments for the partner of the nucleon: a broad,
not-yet discovered partner with a mass of about $1.2$ GeV must be excluded on
the basis of scattering data. The well-established resonance $N(1650)$
provides qualitatively similar results as $N(1535)$ and, in this case, the
theoretical value of the decay width $N(1650)\rightarrow N\eta$ is in
agreement with the experimental one. However, in this scenario it is not clear
how $N(1535)$ fits into the baryonic resonance spectrum. This issue is
discussed in Sec.\ \ref{IVa} presented below. In Sec.\ \ref{IVb} we discuss
the origin of $m_{0}$ in terms of tetraquark and gluon condensates and the
implications for future studies.

\subsection{Outlook 1: enlarged mixing scenario}

\label{IVa}

In this section we briefly describe open problems of the previous results and
present a possible outlook to improve the theoretical description.

A simultaneous description of both resonances $N(1525)$ and $N(1650)$ requires
an extension of the model. In the framework of the mirror assignment, instead
of only two bare nucleon fields $\Psi_{1}$ and $\Psi_{2}$ one should include
two additional bare fields $\Psi_{3}$ and $\Psi_{4}$ with positive and
negative parity, respectively. The latter two are assumed to transform like
$\Psi_{1}$ and $\Psi_{2}$ in Eq.\ (\ref{mirror}). The interesting part of the
enlarged Lagrangian are the bilinear chirally invariant mass terms:
\begin{align}
\mathcal{L}_{\text{mass}}  &  =m_{0}^{(1,2)}\left(  \overline{\Psi}_{2}%
\gamma^{5}\Psi_{1}-\overline{\Psi}_{1}\gamma^{5}\Psi_{2}\right)
+m_{0}^{(3,4)}\left(  \overline{\Psi}_{4}\gamma^{5}\Psi_{3}-\overline{\Psi
}_{3}\gamma^{5}\Psi_{4}\right) \nonumber\\
&  +m_{0}^{(1,4)}\left(  \overline{\Psi}_{4}\gamma^{5}\Psi_{1}-\overline{\Psi
}_{1}\gamma^{5}\Psi_{4}\right)  +m_{0}^{(2,3)}\left(  \overline{\Psi}%
_{2}\gamma^{5}\Psi_{3}-\overline{\Psi}_{3}\gamma^{5}\Psi_{2}\right)  \;.
\end{align}
In the limit $m_{0}^{(1,4)}=m_{0}^{(2,3)}=0$ the bare fields $\Psi_{1}$ and
$\Psi_{2}$ do not mix with the fields $\Psi_{3}$ and $\Psi_{4}.$ The fields
$\Psi_{1}$ and $\Psi_{2}$ generate the states $N(939)$ and $N(1535)$, just as
described in this paper with $m_{0}^{(1,2)}=m_{0}$, while the fields $\Psi
_{3}$ and $\Psi_{4}$ generate the states $N(1440)$ and $N(1650)$, which are
regarded as chiral partners. The term proportional to $m_{0}^{(3,4)}$ induces
a decay of the form $N(1650)\rightarrow N(1440)\pi$ (or $\eta$), but still
$N(1650)$ and $N(1440)$ do not decay into $N\pi(\eta)$.

When in addition the coefficients $m_{0}^{(1,4)}$ and $m_{0}^{(2,3)}$ are
non-zero, a more complicated mixing scenario involving four bare fields
arises. As a consequence, it is possible to account for the decay of both
resonances $N(1550)$ and $N(1650)$ into $N\pi(\eta)$. Moreover, it is well
conceivable that the anomalously small value of the decay width
$N(1550)\rightarrow N\eta$ arises because of destructive interference.
Interestingly, a mixing of bare configurations generating $N(1535)$ and
$N(1650)$ is necessary also at the level of the quark model \cite{isgur}. Note
that in the framework of the generalized mixing scenario, the fields $N(1535)$
and $N(1650)$ are chiral partners of $N(939)$ and $N(1440)$. Due to mixing
phenomena, it is not possible to isolate the chiral partner of the nucleon,
which is present in both resonances $N(1535)$ and $N(1650).$ However, also in
this case the nonzero decay widths of both fields $N(1535)$ and $N(1650)$ are
obtained as a result of non-vanishing $m_{0}$-like parameters.

The mixing scenario outlined above may look at first sight not very useful,
because it involves too many new parameters. However, a quick counting shows
that this is not the case. In addition to the four mass parameters
$m_{0}^{(i,j)}$, we have the already discussed parameters $c_{1}$, $c_{2}$,
$\widehat{g}_{1}$, and $\widehat{g}_{2}$, plus similar parameters $c_{3}$,
$c_{4}$, $\widehat{g}_{3}$, and $\widehat{g}_{4}$ which describe the
interactions of $\Psi_{3,4}$ with mesons. These twelve parameters can be used
to describe the following 14 quantities: the masses of the states $N\equiv
N(939)$, $N(1535)$, $N(1440)$, $N(1650)$, the decay widths $N(1535)\rightarrow
N\pi$, $N(1535)\rightarrow N\eta$, $N(1650)\rightarrow N\pi$,
$N(1650)\rightarrow N\eta$, $N(1440)\rightarrow N\pi$, $N(1440)\rightarrow
N\eta$ (the latter by taking into account the non-zero width of the $N(1440)$
resonance), and the four axial coupling constants $g_{A}^{N}$, $g_{A}%
^{N(1535)}$, $g_{A}^{N(1440)}$, and $g_{A}^{N(1650)}$. A detailed study of
this enlarged scenario, in which the four lightest $J^{P}=\frac{1}{2}^{\pm}$
baryonic resonances are simultaneously included, will be performed in the future.

\subsection{Outlook 2: origin of $m_{0}$}

\label{IVb}

The scattering length $a_{0}^{(+)}$ shows a strong dependence on the mass of
the $\sigma$ meson. A similar situation occurs for $\pi\pi$ scattering at low
energies \cite{denis}. While a light $\sigma$ is favoured by the scattering
data, many other studies show that the $\sigma$ meson -- as the chiral partner
of the pion in the linear sigma model -- should be placed above $1$ GeV and
identified with the resonance $f_{0}(1370)$ rather than the light $f_{0}(600)$
[see Refs.\ \cite{klempt,dynrec} and refs.\ therein]. Indeed, also in the
framework of the linear sigma model used in this paper, the decay width
$f_{0}(600)\rightarrow\pi\pi$ turns out to be too small when the latter is
identified with the chiral partner of the pion \cite{denis}.

When identifying $\sigma$ with $f_{0}(1370)$, two possibilities are left for
$f_{0}(600)$: (i) It is a dynamically generated state arising from the
pion-pion interaction. The remaining scalar states below 1 GeV, $f_{0}(980)$,
$a_{0}(980)$, and $K_{0}^{*}(800)$ can be interpreted similarly. (ii) The
state $f_{0}(600)$ is predominantly composed of a diquark $[u,d]$ (in the
flavor and color antitriplet representation) and an antidiquark $[\overline
{u},\overline{d}]$, i.e., $f_{0}(600)\simeq$ $[\overline{u},\overline
{d}][u,d]$. In this case the light scalar states $f_{0}(600)$, $f_{0}(980)$,
$a_{0}(980)$, and $K_{0}^{*}(800)$ form an additional tetraquark nonet
\cite{jaffeorig,maiani,tq,tqmix,fariborz}. Note that in both cases the
resonance $f_{0}(600)$ -- which is needed to explain $\pi\pi$ and $\pi N$
scattering experiments and also to understand the nucleon-nucleon interaction
potential -- is \emph{not} the chiral partner of the pion. In the following we
concentrate on the implications of scenario (ii) at a qualitative level,
leaving a more detailed study for the future. First, a short digression on the
dilaton field is necessary.

Dilatation invariance of the QCD Lagrangian in the chiral limit is broken by
quantum effects. This situation can be taken into account in the framework of
a chiral model by introducing the dilaton field $G$ \cite{salomone}. The
corresponding dilaton potential reflects the trace anomaly of QCD as
underlying theory and has the form $V(G)\propto G^{4}\left(  \log\frac
{G}{\Lambda_{G}}-\frac{1}{4}\right)  $, where $\Lambda_{G}\sim\Lambda_{QCD}$
is the only dimensional quantity which appears in the full effective
Lagrangian in the chiral limit. Due to the non-zero expectation value of $G$,
a shift $G\rightarrow G_{0}+G$ is necessary: the fluctuations around the
minimum correspond to the scalar glueball, whose mass is placed at $M_{G}%
\sim1.7$ GeV by lattice QCD calculations \cite{lattglue} and by various
phenomenological studies \cite{gluephen}. (Beyond the chiral limit, also the
parameter $h_{0}$ in Eq.\ (\ref{meslag}), which describes explicit symmetry
breaking due to the non-zero valence quark masses, appears as an additional
dimensionful quantity.)

We assume that, in the chiral limit, the full interaction potential
$V(\Phi,L_{\mu},R_{\mu},\Psi_{1},\Psi_{2},G,\chi)$ is dilatation invariant up
to the term $\propto\log\frac{G}{\Lambda_{G}}$ and that it is finite for any
finite value of the fields, i.e., only terms of the kind $G^{2}\mathrm{Tr}%
\left[  \Phi^{\dagger}\Phi\right]  $, $\mathrm{Tr}\left[  \Phi^{\dagger}%
\Phi\right]  ^{2}\, , \ldots$ are retained. By performing the shift
$G\rightarrow G_{0}+G$, the term $G^{2}\mathrm{Tr}\left[  \Phi^{\dagger}%
\Phi\right]  $ becomes $G_{0}^{2}\mathrm{Tr}\left[  \Phi^{\dagger}\Phi\right]
+ \ldots$, where the dots refer to glueball-meson interactions. Identifying
$\mu^{2} \sim G_{0}^{2}$, a term $G_{0}^{2}\mathrm{Tr} \left[  \Phi^{\dagger
}\Phi\right]  $ is already present in our Lagrangian (\ref{meslag}), but the
glueball-hadron interactions are neglected. Note that a term of the kind
$G^{-4}\mathrm{Tr}\left[  \partial_{\mu}\Phi^{\dagger} \partial^{\mu}%
\Phi\right]  ^{2}$ is not allowed because of our assumption that the potential
is finite. Following this line of arguments, our Lagrangian (\ref{meslag})
cannot contain operators of order higher than four \cite{dynrec}, because such
operators must be generated from terms with inverse powers of $G$. E.g., upon
shifting $G$, the above mentioned term would generate an order-eight operator
of the kind $G_{0}^{-4}\mathrm{Tr}\left[  \partial_{\mu}\Phi^{\dagger}
\partial^{\mu}\Phi\right]  ^{2}$.

Let us now turn to the mass term $\sim m_{0}$ in Eq.\ (\ref{nucl lagra}),
\begin{equation}
\label{massterm}m_{0}(\overline{\Psi}_{1L}\Psi_{2R} -\overline{\Psi}_{1R}%
\Psi_{2L} -\overline{\Psi}_{2L}\Psi_{1R} +\overline{\Psi}_{2R}\Psi_{1L})\;.
\end{equation}
The parameter $m_{0}$ has the dimension of mass and is the only term in the
baryon sector, which is not dilatation invariant. In order to render it
dilatation invariant while simultaneously preserving chiral symmetry, we can
couple it to the chirally invariant dilaton field $G$. Moreover, in the
framework of $U(2)_{R}\times U(2)_{L}$ chiral symmetry also the above
mentioned tetraquark field, denoted as $\chi\equiv\frac{1}{2} \lbrack
\overline{u},\overline{d}][u,d]$, is invariant under chiral transformations.
We then write the following dilatation-invariant interaction term:
\begin{equation}
\label{tqbar}\left(  a\chi+bG\right)  \, (\overline{\Psi}_{1L}\Psi
_{2R}-\overline{\Psi}_{1R}\Psi_{2L} -\overline{\Psi}_{2L}\Psi_{1R}%
+\overline{\Psi}_{2R}\Psi_{1L})\;,
\end{equation}
where $a$ and $b$ are dimensionless coupling constants.

When shifting both fields around their vacuum expectation values
$\chi\rightarrow\chi_{0}+\chi$ and $G\rightarrow G_{0}+G$ we recover the term
(\ref{massterm}) of our Lagrangian by identifying
\begin{equation}
m_{0}=a\chi_{0}+bG_{0}\;,
\end{equation}
where $\chi_{0}$ and $G_{0}$ are the tetraquark and gluon condensates, respectively.

Note that the present discussion holds true also in the highly excited part of
the baryon sector: as described in Ref.\ \cite{Glozman,cohen}, the heavier the
baryons, the less important becomes the quark condensate $\varphi$: For two
heavy chiral partners $B$ and $B^{*}$, one expects a mass degeneracy of the
form $m_{B} \simeq m_{B^{\ast}}\simeq m_{0}$. We expect the gluon condensate
$G_{0}$ to be the dominant term in this sector, $m_{0} \simeq b G_{0}$. In
fact, the tetraquark condensate is also related to the chiral condensate in
the vacuum \cite{tqmix,Heinz:2008cv} and -- while potentially important for
low-lying states like the nucleon and its partner -- its role should also
diminish when considering very heavy baryons.

We now return to the nucleon and its partner and concentrate on their
interaction with the tetraquark field $\chi$. From the point of low-energy
phenomenology, the tetraquark field $\chi$ is very interesting because the
corresponding excitation is expected to be lighter than the gluonium and the
scalar quarkonium states, for instance $m_{\chi}\sim M_{f_{0}(600)}\sim0.6$
GeV. A nucleon-tetraquark interaction of the kind $a\chi(\overline{\Psi}%
_{1L}\Psi_{2R} -\overline{\Psi}_{1R}\Psi_{2L} -\overline{\Psi}_{2L}\Psi_{1R}
+\overline{\Psi}_{2R}\Psi_{1L})$ arising from Eq.\ (\ref{tqbar}) would then
contribute to pion-pion and nucleon-pion scattering and possibly improve the
agreement with experimental data.

Moreover, there is also another interesting consequence: in virtue of
Eq.\ (\ref{tqbar}) the state $\chi$ appears as intermediate state in
nucleon-nucleon interactions and, due to its small mass, is likely to play an
important role in the one-meson exchange picture for the nucleon-nucleon
potential. This raises the interesting question whether a \emph{tetraquark\/}
is the scalar state which mediates the middle-range attraction among nucleons,
in contrast to the standard picture where this task is performed by a
quark-antiquark state. Let us further elucidate this picture by a simple and
intuitive example. Let us consider the nucleon as a quark-diquark bound state.
The standard picture of one-boson exchange in the nucleon-nucleon interaction
consists of exchanging the two quarks between the nucleons. However, one could
well imagine that instead of the quarks one exchanges the two diquarks between
the nucleons. Note that these diquarks are in the correct color and flavor
antitriplet representations in order to form a tetraquark of the type
suggested by Jaffe \cite{jaffeorig}, such as the meson $\chi$ discussed here.
A full analysis must include a detailed study of mixing between all scalar states.

As a last subject we discuss how the nucleon mass might evolve at non-zero
temperature and density. In particular, in the high-density region of the
so-called ``quarkyonic phase'' \cite{quarkyonic} hadrons are confined but
chiral symmetry is (almost) restored, i.e., the chiral condensate
(approximately) vanishes. What are the properties of the nucleon in this
phase? In the framework of the Lagrangian (\ref{nucl lagra}), when
$\varphi\rightarrow0$, the masses of both the nucleon and its partner approach
a constant value $m_{0}$. Then, the first naive answer is that we expect a
nucleon mass of about $500$ MeV in this phase. The situation is, however, more
complicated than this. In fact, as discussed in this section the term $m_{0}$
is not simply a constant but is related to other condensates. The behavior of
these condensates at non-zero $T$ and $\mu$ is then crucial for the
determination of the nucleon mass. Interestingly, in Ref.\ \cite{Heinz:2008cv}
it is shown that the tetraquark condensate does \emph{not} vanish but rather
increases for increasing $T$. A future study at non-zero $T$ and $\mu$ must
include both the tetraquark and the gluon condensate in the same framework.

\section*{Acknowledgements}

The authors thank L.\ Glozman, T.\ Kunihiro, S.\ Leupold, D.\ Parganlija,
R.\ Pisarski, and T.\ Takahashi for useful discussions. The work of S.G.\ was
supported by GSI Darmstadt under the F\&E program. The work of F.G.\ was
partially supported by BMBF. The work of D.H.R.\ was supported by the ExtreMe
Matter Institute EMMI. This work was (financially) supported by the Helmholtz
International Center for FAIR within the framework of the LOEWE program
launched by the State of Hesse.

\appendix

\section{Vector-meson self-interactions}

\label{appendixC}

In this appendix, we present the terms $\mathcal{L}_{3}$ and $\mathcal{L}_{4}$
of Eq.\ (\ref{meslag}):
\begin{align}
\mathcal{L}_{3}  &  =-2ig_{2}\left(  \mathrm{Tr}\{L_{\mu\nu}[L^{\mu},L^{\nu
}]\} +\mathrm{Tr}\{R_{\mu\nu}[R^{\mu},R^{\nu}]\}\right) \nonumber\\
&  -2g_{3}\left(  \mathrm{Tr}[\left(  \partial_{\mu}L_{\nu}+\partial_{\nu
}L_{\mu}\right)  \{L^{\mu},L^{\nu}\}]+\mathrm{Tr}[\left(  \partial_{\mu}%
R_{\nu}+\partial_{\nu}R_{\mu}\right)  \{R^{\mu},R^{\nu}\}]\right)  ,\;
\end{align}
and
\begin{align}
\mathcal{L}_{4}  &  =g_{4}\left\{  \mathrm{Tr}\left[  L^{\mu}L^{\nu}L_{\mu
}L_{\nu}\right]  +\mathrm{Tr}\left[  R^{\mu}R^{\nu}R_{\mu}R_{\nu}\right]
\right\}  +g_{5}\left\{  \mathrm{Tr}\left[  L^{\mu}L_{\mu}L^{\nu}L_{\nu
}\right]  +\mathrm{Tr}\left[  R^{\mu}R_{\mu}R^{\nu}R_{\nu}\right]  \right\}
\nonumber\\
&  +g_{6}\mathrm{Tr}\left[  R^{\mu}R_{\mu}\right]  \, \mathrm{Tr}\left[
L^{\nu}L_{\nu}\right]  +g_{7}\left\{  \mathrm{Tr}[L^{\mu}L_{\mu}]\,
\mathrm{Tr}[L^{\nu}L_{\nu}] +\mathrm{Tr}[R^{\mu}R_{\mu}]\, \mathrm{Tr}[R^{\nu
}R_{\nu}]\right\}  \; .
\end{align}
The coupling constants $g_{k}$ with $k=2,\ldots,7$ are not relevant for the
present work.

\section{Meson sector}

\label{appendixA}

In the mesonic Lagrangian (\ref{meslag}), there are ten parameters:
$\lambda_{1}$, $\lambda_{2}$, $c$, $h_{0}$, $h_{1}$, $h_{2}$, $h_{3}$,
$\mu^{2}$, $g_{1}$, and $m_{1}$. In the following, we describe how to relate
them to the physical meson masses and the pion decay constant.

If chiral symmetry is spontaneously broken, the scalar-isoscalar field
$\sigma$ develops a non-vanishing vacuum expectation value (v.e.v.),
$\langle\sigma\rangle\equiv\varphi$, the so-called chiral condensate. In order
to proceed, we have to shift $\sigma$ by its v.e.v., $\sigma\rightarrow
\varphi+\sigma$. The chiral condensate is identified with the minimum of the
potential energy density $V(\varphi)$, cf.\ Eq.\ (\ref{meslag}):
\begin{align}
V(\varphi)  &  =\frac{1}{2}(\mu^{2}-c)\varphi^{2} +\frac{1}{4}\left(
\lambda_{1}+\frac{\lambda_{2}}{2}\right)  \varphi^{4}-h_{0}\varphi\; ,\\
0  &  =\frac{dV}{d\varphi}= \left[  \mu^{2}-c +\left(  \lambda_{1}%
+\frac{\lambda_{2}}{2}\right)  \varphi^{2}\right]  \varphi-h_{0}\;.
\end{align}
After the shift $\sigma\rightarrow\varphi+\sigma$ a mixing term between
axial-vector and pseudoscalar mesons arises; for instance between $a_{1}%
$-meson and pion it is of the form $-g_{1} \vec{a}^{\mu}_{1} \cdot
\partial_{\mu}\vec{\pi}$. The standard way to treat this term is to eliminate
it by a shift of the axial-vector fields. Then, in order to recover the
canonical normalization of the pseudoscalar fields, one has to introduce a
corresponding wavefunction renormalization factor. For $a_{1}$-meson and pion
this operation has the form
\begin{equation}
\vec{a}^{\mu}_{1}\rightarrow\vec{a}^{\mu}_{1}+Zw\,\partial^{\mu} \vec{\pi}\;,
\;\; \vec{\pi}\rightarrow Z\vec{\pi}\;, \;\;\;\; \mathrm{where}\;\;\;
w=\frac{g_{1}\varphi}{m_{a_{1}}^{2}}\;,\;\; Z^{2} =\frac{m_{a_{1}}^{2}%
}{m_{a_{1}}^{2}-(g_{1}\varphi)^{2}}\;. \label{fieldtr}%
\end{equation}
The meson masses are then given by:
\begin{align}
m_{\sigma}^{2}  &  = \mu^{2}-c + 3 \left(  \lambda_{1}+\frac{\lambda_{2}}%
{2}\right)  \varphi^{2}\;, \;\; m_{a_{0}}^{2}=\mu^{2}+c + \left(  \lambda
_{1}+3\frac{\lambda_{2}}{2}\right)  \varphi^{2} \;,\\
m_{\eta_{N}}^{2}  &  =Z^{2} \left[  \mu^{2}+c +\left(  \lambda_{1}%
+\frac{\lambda_{2}}{2}\right)  \varphi^{2}\right]  \; , \;\; m_{\pi}^{2}%
=Z^{2}\left[  \mu^{2}-c +\left(  \lambda_{1}+\frac{\lambda_{2}}{2}\right)
\varphi^{2}\right]  =\frac{Z^{2}h_{0}}{\varphi}\;,\\
m_{\omega}^{2}  &  =m_{\rho}^{2} =m_{1}^{2}+\frac{\varphi^{2}}{2}(h_{1}%
+h_{2}+h_{3})\; ,\;\; m_{f_{1}}^{2}=m_{a_{1}}^{2}=m_{1}^{2}+(g_{1}\varphi)^{2}
+\frac{\varphi^{2}}{2}(h_{1}+h_{2}-h_{3})\;.
\end{align}
Note that only the linear combination $h_{1}+h_{2}$ enters these equations, so
only nine out of the original ten parameters are determined by the meson
masses. However, in the following considerations, only the sum $h_{1}+h_{2}$
will enter, so we do not need to determine $h_{1}$ and $h_{2}$ independently.
We therefore have six physical meson masses in order to determine nine
parameters. A seventh physical quantity is the pion decay constant, $f_{\pi}$,
which we determine from the axial current, $J_{A\mu}^{a}=\frac{\varphi}
{Z}\partial_{\mu}\pi^{a}+\ldots\equiv f_{\pi}\partial_{\mu}\pi^{a} + \ldots$,
i.e., $\varphi=Zf_{\pi}$.

This leaves us with two independent parameters, which turn out to be $g_{1}$
and $m_{1}$. The latter only enters the isospin-even pion-nucleon scattering
length. We shall leave it as a free parameter to study the dependence of
$a_{0}^{(+)}$ on $m_{1}$.

For the sake of convenience, we shall replace the coupling constant $g_{1}$ by
the pseudoscalar wavefunction renormalization factor $Z$. This is achieved
with the help of the relation (\ref{fieldtr}),
\begin{equation}
g_{1}(Z)=\frac{m_{a_{1}}}{Zf_{\pi}}\sqrt{1-\frac{1}{Z^{2}}}\;. \label{g1}%
\end{equation}
In this work we fix $m_{a_{1}}=1.23$ GeV which is the central value quoted in
Ref.\ \cite{PDG}. In fact, it is technically easier to use $Z$ than $g_{1}$ as
independent parameter: while $g_{1}$ is a unique function of $Z$, the function
$Z(g_{1})$ would be multi-valued.

It remains to determine $Z$. For this purpose we use the decay width
$a_{1}\rightarrow\pi\gamma$. The experimental value quoted by the PDG is
$\Gamma_{a_{1}\rightarrow\pi\gamma}^{\exp}=(640\pm246)$ keV \cite{PDG}. The
theoretical expression is obtained by minimal coupling of the photon in the
meson sector and only depends on $Z$:
\begin{equation}
\Gamma_{a_{1}\rightarrow\pi\gamma}=\frac{\alpha}{24}\,m_{a_{1}}\,(Z^{2}%
-1)\left(  1-\frac{m_{\pi}^{2}}{m_{a_{1}}^{2}}\right)  ^{3}\;, \label{a1decay}%
\end{equation}
where $\alpha=1/137.$ Using the experimental value quoted above we derive
$Z=1.67\pm0.19$. The quantities $m_{1}$ and $Z$ are the only independent
parameters from the mesonic sector, which enter the determination of the axial
coupling constants of the nucleon and its chiral partner, the decay widths for
$N^{\ast}\rightarrow NP$, and the $N\pi$ scattering lengths. Due to the large
uncertainty, we shall employ Eq.\ (\ref{a1decay}) together with the
constraints from the baryon sector, cf.\ Sec.\ \ref{III}, to perform a
simultaneous fit of all relevant parameters in the baryonic sector, i.e.,
$c_{1},\,c_{2},\,Z,$ and $m_{0}$.

It should be noted that the inclusion of the axial-vector degrees of freedom
is the ultimate reason which allows for a correct determination of the
axial-coupling constants $g_{A}^{N}$ and $g_{A}^{N\ast}$. We can easily
convince ourselves of this fact by assuming the contrary, i.e., studying the
case where the axial-vector mesons are absent. This can be achieved either by
setting $g_{1}$ to zero, or by sending the $a_{1}$ mass to infinity. In both
cases, $Z=\left[  1-\left(  g_{1}\varphi/m_{a_{1}}\right)  ^{2}\right]
^{-1/2} \rightarrow1 + O[(g_{1} \varphi/m_{a_{1}})^{2}]$. Then, from
Eq.\ (\ref{gas}), we obtain $g_{A}^{(1)}= - g_{A}^{(2)} = 1$, and the physical
axial coupling constants are $g_{A}^{(N)} = - g_{A}^{(N^{\ast})} =\tanh
\delta\leq1$, in contradiction to the experimental values.

The next question is, whether the experimental value of the $a_{1}$ mass is
not too large compared to the `natural scale' of the problem, so that the
correct description of the axial-coupling constants is impossible. The natural
scale is given by the scale of chiral symmetry breaking, i.e., by the value of
$\varphi$, possibly multiplied by a constant of order one. If we take the
natural scale to be $g_{1}\varphi\sim g_{1}f_{\pi}\simeq600$ MeV, then indeed
$g_{1}\varphi/m_{a_{1}}\sim1$, i.e., the $a_{1}$ mass is not too large
compared to the natural scale of the problem. This can also be seen from the
fact that the $a_{1}\rightarrow\pi\gamma$ decay requires $Z\simeq1.67>1$,
i.e., $g_{1}\varphi$ must be of order $m_{a_{1}}$. If $m_{a_{1}}$ were large,
a fit of $g_{A}^{(N^{\ast})}$ to the lattice data would lead to an unnaturally
large $c_{2}$. But this problem does not emerge because $m_{a_{1}}$ is not
large when compared to the natural scale of the model.

\section{Details of the calculations}

\label{appendixB}

\subsection{Axial coupling constants}

From the baryonic Lagrangian (\ref{nucl lagra}) we select the terms which are
relevant for the derivation of the baryonic axial coupling constants:
\[
\mathcal{L}^{ax}=i\overline{\Psi}_{1}\gamma^{\mu}\partial_{\mu}\Psi
_{1}+i\overline{\Psi}_{2}\gamma^{\mu}\partial_{\mu}\Psi_{2}-c_{1}%
\overline{\Psi}_{1}\gamma^{\mu}\gamma^{5}\vec{t}\cdot\vec{a}_{1\mu}\Psi
_{1}+c_{2}\overline{\Psi}_{2}\gamma^{\mu}\gamma^{5}\vec{t}\cdot\vec{a}_{1\mu
}\Psi_{2}\;,
\]
in which the interactions of $\Psi_{1}$ and $\Psi_{2}$ with the $a_{1}$-meson
are retained. After performing the shift of the axial field $\vec{a}_{1}^{\mu
}\rightarrow\vec{a}_{1}^{\mu}+Zw\,\partial^{\mu}\vec{\pi}$ we obtain:
\begin{equation}
\mathcal{L}^{ax}=i\overline{\Psi}_{1}\gamma^{\mu}\partial_{\mu}\Psi
_{1}+i\overline{\Psi}_{2}\gamma^{\mu}\partial_{\mu}\Psi_{2}-Zwc_{1}%
\overline{\Psi}_{1}\gamma^{\mu}\gamma^{5}\vec{t}\cdot\partial_{\mu}\vec{\pi
}\Psi_{1}+Zwc_{2}\overline{\Psi}_{2}\gamma^{\mu}\gamma^{5}\vec{t}\cdot
\partial_{\mu}\vec{\pi}\Psi_{2}+\ldots\;.
\end{equation}
The axial current is calculated as
\begin{equation}
\mathcal{J}_{A\mu}^{i}=\frac{\partial\mathcal{L}}{\partial(\partial^{\mu}%
\Psi_{1})}(\delta\Psi_{1})^{i}+\frac{\partial\mathcal{L}}{\partial
(\partial^{\mu}\Psi_{2})}(\delta\Psi_{2})^{i}+\frac{\partial\mathcal{L}%
}{\partial(\partial^{\mu}\pi^{j})}(\delta\pi^{j})^{i}\;,
\end{equation}
where $(\delta\Psi_{1})^{i}=i\gamma^{5}t^{i}\Psi_{1}$, $(\delta\Psi_{2}%
)^{i}=-i\gamma^{5}t^{i}\Psi_{2}$, $(\delta\pi^{j})^{i}=\delta^{ij}%
(\sigma+\varphi)/Z$.

We obtain:
\begin{equation}
\mathcal{J}_{A\mu}^{i}=g_{A}^{(1)}\overline{\Psi}_{1}\gamma_{\mu}\gamma
^{5}t^{i}\Psi_{1}+g_{A}^{(2)}\overline{\Psi}_{2}\gamma_{\mu}\gamma^{5}%
t^{i}\Psi_{2}+\ldots\;,
\end{equation}
where, taking into account that $w=(1-Z^{-2})/(g_{1}\varphi)$:
\begin{equation}
g_{A}^{(1)}=1-\varphi wc_{1}=1-\frac{c_{1}}{g_{1}}\left(  1-\frac{1}{Z^{2}%
}\right)  \;,\;\;\;g_{A}^{(2)}=-1+\frac{c_{2}}{g_{1}}\left(  1-\frac{1}{Z^{2}%
}\right)  \;,
\end{equation}
which are Eqs.\ (\ref{gas}).

In order to obtain the axial coupling constants of the physical fields, we
make use of Eq.\ (\ref{mixing}):
\begin{equation}
\mathcal{J}_{A \mu}^{i}=g_{A}^{N}\overline{N} \gamma_{\mu}\gamma^{5}t^{i}
Ne^{\delta} +g_{A}^{N^{\ast}}\overline{N}^{\ast} \gamma_{\mu}\gamma^{5}t^{i}
N^{\ast}e^{-\delta}+\ldots
\end{equation}
where
\begin{equation}
g_{A}^{N}=\frac{1}{2\cosh\delta} \left(  e^{\delta}\,g_{A}^{(1)}+e^{-\delta
}\,g_{A}^{(2)}\right)  \;,\;\; g_{A}^{N^{\ast}}=\frac{1}{2\cosh\delta} \left(
e^{-\delta}\,g_{A}^{(1)}+e^{\delta}\,g_{A}^{(2)}\right)  \;,
\end{equation}
which are Eqs.\ (\ref{axialcoupl}).

\subsection{Decay widths}

After the field transformation $\sigma\rightarrow\varphi+ \sigma$ and
(\ref{fieldtr}) discussed in Appendix \ref{appendixA} have been performed, we
isolate the terms relevant for the decay $N^{\ast}\rightarrow NP$. In the
following, we only discuss $P=\pi^{0},\,\eta_{N}$, the other isospin
components can be obtained similarly:
\begin{equation}
\mathcal{L}_{NN^{\ast}P}=iA\overline{N}^{\ast}NP+B\overline{N}^{\ast}
\gamma^{\mu}N\partial_{\mu}P-iA\overline{N}N^{\ast}P+B\overline{N}\gamma^{\mu
}N^{\ast}\partial_{\mu}P\;. \label{lnp}%
\end{equation}
where
\begin{equation}
A=-\frac{Z(\widehat{g}_{1}-\widehat{g}_{2})}{4\cosh\delta}\;, \;\;\;B=-\frac
{Zw(c_{1}+c_{2})}{4\cosh\delta}\;.
\end{equation}
The decay amplitude for the process $N^{\ast}\rightarrow NP$ reads:
\begin{equation}
-iM_{\alpha\beta}=i\overline{u}^{N}_{\beta}(\vec{k}_{1}) C\, u^{N^{\ast}%
}_{\alpha}(\vec{k}=0)\;,
\end{equation}
where $C=-iA+iB\gamma_{\rho} k_{2}^{\rho}$. Averaging over initial states and
summing over final states, we obtain the following squared amplitude:
\begin{equation}
\overline{\left\vert -iM_{N^{\ast}\rightarrow NP}\right\vert ^{2}}=\frac{1}%
{2}\sum_{\alpha,\beta} \left\vert -iM_{\alpha\beta}\right\vert ^{2}=\frac
{1}{2} \sum_{\alpha,\beta}\left[  \overline{u}^{N}_{\beta}(\vec{k}_{1}) C\,
u^{N^{\ast}}_{\alpha} (\vec{k}=0)\right]  \left[  \overline{u}^{N^{\ast}%
}_{\alpha} (\vec{k}=0)C^{\prime}\,u^{N}_{\beta}(\vec{k}_{1})\right]  \;,
\end{equation}
with $C^{\prime}=iA-iB\gamma_{\rho}k_{2}^{\rho}=-C$. Using the well-known
properties of the traces of $\gamma$ matrices leads to the result:
\begin{align}
\overline{\left\vert -iM\right\vert ^{2}}  &  =\frac{1}{2}\sum_{\alpha,\beta}
\left\vert -iM_{\alpha\beta}\right\vert ^{2} =\frac{1}{2}\mathrm{Tr}\left[
C\,\frac{\gamma^{\mu}k_{\mu}+m_{N^{\ast}}}{2m_{N^{\ast}}}\, C^{^{\prime}}\,
\frac{\gamma^{\nu}k_{1,\nu}+m_{N}}{2m_{N}}\right] \nonumber\\
&  =\frac{A^{2}}{2}\mathrm{Tr}\left[  \frac{\gamma^{\mu}k_{\mu}+m_{N^{\ast}}%
}{2m_{N^{\ast}}}\, \frac{\gamma^{\mu}k_{1,\mu}+m_{N}}{2m_{N}}\right]
+\frac{B^{2}}{2}\mathrm{Tr}\left[  \gamma_{\rho}k_{2}^{\rho}\, \frac
{\gamma^{\mu}k_{\mu}+m_{N^{\ast}}}{2m_{N^{\ast}}}\, \gamma_{\rho}k_{2}^{\rho
}\, \frac{\gamma^{\mu}k_{1,\mu}+m_{N}}{2m_{N}}\right] \nonumber\\
&  -AB\,\mathrm{Tr}\left[  \frac{\gamma^{\mu}k_{\mu}+m_{N^{\ast}}}%
{2m_{N^{\ast}}}\, \gamma_{\rho}k_{2}^{\rho}\, \frac{\gamma^{\mu}k_{1,\mu
}+m_{N}}{2m_{N}}\right] \nonumber\\
&  =\frac{A^{2}}{2}\left(  \frac{E_{N}}{m_{N}}+1\right)  +\frac{B^{2}}%
{2}\left[  \left(  m_{N^{\ast}}^{2}-m_{N}^{2}-m_{P}^{2}\right)  \, \frac
{E_{P}}{m_{N}} +m_{P}^{2}\left(  1-\frac{E_{N}}{m_{N}}\right)  \right]
\nonumber\\
&  -AB\left(  \frac{m_{N^{\ast}}^{2}-m_{N}^{2}-m_{P}^{2}}{2m_{N}}+E_{P}
\right)  \; .
\end{align}
The full decay width is obtained including all isospin states for the pion and
by replacing the unphysical state $\eta_{N}$ with the physical $\eta$ meson.
The result is
\begin{equation}
\Gamma_{N^{\ast}\rightarrow NP}=\lambda_{P} \frac{k_{P}}{2\pi}\frac{m_{N}%
}{m_{N^{\ast}}} \overline{\left\vert -iM_{N^{\ast}\rightarrow NP}\right\vert
^{2}}\;,
\end{equation}
which leads to Eq.\ (\ref{npidecay}).

\end{document}